# Symbiotic Blockchain Consensus: Cognitive Backscatter Communications-enabled Wireless Blockchain Consensus

Haoxiang Luo, Qianqian Zhang, Gang Sun, *Senior Member, IEEE*,
Hongfang Yu, *Member, IEEE*, and Dusit Niyato, *Fellow, IEEE*

*Abstract*—The wireless blockchain network (WBN) concept, born from the blockchain deployed in wireless networks, has appealed to many network scenarios. Blockchain consensus mechanisms (CMs) are key to enabling nodes in a wireless network to achieve consistency without any trusted entity. However, consensus reliability will be seriously affected by the instability of communication links in wireless networks. Meanwhile, it is difficult for nodes in wireless scenarios to obtain a timely energy supply. Energy-intensive blockchain functions can quickly drain the power of nodes, thus degrading consensus performance. Fortunately, a symbiotic radio (SR) system enabled by cognitive backscatter communications can solve the above problems. In SR, the secondary transmitter (STx) transmits messages over the radio frequency (RF) signal emitted from a primary transmitter (PTx) with extremely low energy consumption, and the STx can provide multipath gain to the PTx in return. Such an approach is useful for almost all vote-based CMs, such as the Practical Byzantine Fault-tolerant (PBFT)-like and the RAFT-like CMs. This paper proposes symbiotic blockchain consensus (SBC) by transforming 6 PBFT-like and 4 RAFT-like state-of-the-art (SOTA) CMs to demonstrate universality. These new CMs will benefit from mutualistic transmission relationships in SR, making full use of the limited spectrum resources in WBN. Simulation results show that SBC can increase the consensus success rate of PBFT-like and RAFT-like by 54.1% and 5.8%, respectively, and reduce energy consumption by 9.2% and 23.7%, respectively.

This work was supported by the Natural Science Foundation of Sichuan Province under Grant 2022NSFSC0913. An earlier version of this paper was presented in part at the IEEE Global Communications Conference (GLOBECOM) 2023 [1]. *(Corresponding author: Gang Sun.)*

H. Luo and G. Sun are with the Key Laboratory of Optical Fiber Sensing and Communications (Ministry of Education), University of Electronic Science and Technology of China, Chengdu 611731, China (e-mail: lhx991115@163.com; gangsun@uestc.edu.cn).
Q. Zhang is with the National Key Laboratory of Wireless Communications, University of Electronic Science and Technology of China, Chengdu 611731, China (e-mail: qqzhang_kite @163.com).
H. Yu is with the Key Laboratory of Optical Fiber Sensing and Communications (Ministry of Education), University of Electronic Science and Technology of China, Chengdu 611731, China, and also with Peng Cheng Laboratory, Shenzhen 518066, China (e-mail: yuhf@uestc.edu.cn).
D. Niyato is with the School of Computer Science and Engineering, Nanyang Technological University, Singapore 639798 (e-mail: dniyato@ntu.edu.sg).

## I. INTRODUCTION

BLOCKCHAIN, as a revolutionary distributed ledger technology, has outstanding advantages such as immobility, and decentralization, which is believed to be expected to change the information interaction mode of our future society [1-2]. In recent years, due to the above advantages, it has also been widely used in various network scenarios, such as the Internet of Things (IoT) [3], Internet of Vehicles (IoV) [4], Internet of Energy (IoE) [5], Decentralized Autonomous Organizations (DAOs) [6], etc. Meanwhile, in the field of wireless networks and communications, blockchain is also considered to be a disruptive potential technology in 6G communications [7-8].

The CM in the blockchain is key to the proper functioning of the WBN, which allows nodes in the network to establish a trust foundation without any trusted third party. It plays an important role in achieving the consistency of distributed systems [9]. Recently, the typical voting-based CM, Practical Byzantine Fault Tolerance (PBFT) [10] and RAFT [11] have been highlighted by the WBN for their high fault tolerance and clear workflow. They are also typical representatives of the consortium and private chains [12], and have been widely researched. For example, Onireti *et al.* [13] have proposed a minimum number of replicas to maintain the consensus activity of the wireless PBFT CM. In [14], Li *et al.* have studied the consensus performance of wireless RAFT CM by the Markov chain and probability theory. Moreover, Luo *et al.* [15-16] have studied the performance of wireless PBFT and RAFT consensus with terahertz and millimeter wave signals in 6G communications. In [4], the authors have investigated the performance of wireless PBFT and RAFT consensus in a wireless IoV environment.

### A. Research Motivation

Although there have been a lot of works focused on the PBFT and RAFT performances in wireless networks, these analysis results all point to a worrying fact. Unstable communication links in wireless networks can seriously affect the consensus success rate of PBFT and RAFT [17-18], which relies on multiple rounds of communication to achieve consistency among nodes in the network. Moreover, there are multiple communication processes in WBN at the same time, and interference will occur among these communication links, which is a problem rarely considered in the existing work. In addition, in wireless networks, the nodes participating in the consensus find it difficult to obtain a timely power supply.

And blockchain happens to be an energy-intensive function. As a result, nodes in a WBN are likely to quickly run out of power and go offline, reducing consensus performance [1]. According to the above description, the **low consensus success rate** caused by unstable communication links and signal interference in wireless networks, and **high energy consumption** are the major challenges currently limiting the further development of wireless blockchain consensus.

To alleviate the above two problems, some efforts have been made to better migrate PBFT and RAFT consensus to wireless networks. For instance, [19] and [20] have respectively optimized the consensus processes of PBFT and RAFT to achieve higher consensus efficiency in the WBN. Meanwhile, in [17], the authors have designed a fast and highly fault-tolerant communication protocol for WBNs. However, these methods do not touch the core problem, that is, to enhance the communication links in the WBN under the circumstance of limited energy and spectrum resources, to guarantee consensus performance. Moreover, with the advent of B5G/6G communications, the communication equipment and systems added to the radio will continue to increase, which will further limit the spectrum resources of WBNs. The consensus performance of the WBN largely depends on the spectrum resources of wireless links, and the allocation of the limited spectrum resources in the wireless scenario is often random [21]. Therefore, the failure to fully exploit spectrum resources in the WBN to enhance consensus performance is fundamentally an incurable method.

Fortunately, symbiotic radio (SR) based on backscatter communication [22], as another potential 6G enabling technology, provides support for the actual deployment of PBFT and RAFT CMs in wireless scenarios for efficient and reliable WBNs. An SR comprises a primary communication system, consisting of a PTx and a PRx, along with a secondary communication system, composed of an STx and an SRx. Through appropriate deployment, the two communication systems will form a mutualistic transmission relationship within the SR [23]. This mutualistic transmission mechanism reveals that the primary system can benefit from multipath gain facilitated by the secondary system, while the secondary system can achieve extreme-low-power [1] (with microwatt magnitude) backscatter transmission by leveraging the RF signal from the primary system [22-24]. The advantages derived from the primary and secondary systems offer effective solutions to address the challenges of transmission reliability and energy consumption in the WBN. Furthermore, the utilization of SR enables cost-efficient servicing of the WBN by implementing a backscatter circuit with two load impedances from the nodes within the secondary system [23].

Consequently, the SR emerges as a pivotal approach in WBNs, enabling the optimal utilization of spectrum resources and offering a fundamental solution to address the challenges of unreliable communication links and high energy consumption in wireless CMs.

### B. Our Contributions

Based on the above benefits of SR, we intend to deploy

---

[1] Backscatter communication is considered energy-free because of the large power magnitude difference between active communication and itself [22-24].

---

cognitive backscatter communication in WBNs to solve the inherent problems of unreliable transmission and high energy consumption. It innovatively proposes the concept of the SBC. To the best of our knowledge, **this is the first work to introduce SR into WBNs**. The contributions are as follows

- We propose an **SBC** and transform **6 PBFT-like and 4 RAFT-like SOTA CMs** as representatives to prove its universality. This approach can enhance some communication signal strength and replace some communication processes with cognitive backscatter communications in CMs at almost negligible cost.
- We investigate **how to deploy SR systems in the WBN to implement the SBC**, determining which nodes act as the primary system at what time, and which nodes act as the secondary system at what time. To avoid interference between communication links, SBC adopts frequency division (FD) or time division (TD) multiplexing based on available spectrum resources. We demonstrate the specific workflows of symbiotic PBFT (S-PBFT) and symbiotic RAFT (S-RAFT) as examples.
- We theoretically analyze **how cognitive backscatter communication benefits the SBC**. Specifically, we derive the consensus success rate, energy consumption, consensus latency, and communication overhead of S-PBFT and S-RAFT for both FD and TD scenarios.
- Through numerical simulation with 10 aforementioned SOTA CMs, we show that SBC can not only improve the consensus success rate and reduce energy consumption, but also optimize the consensus latency, and communication overhead. Additionally, the results provide a comprehensive evaluation of these CMs in a wireless environment.

### C. Structure of This Paper

The rest of the contents are arranged as follows. Section II is the related work. Section III introduces the PBFT and RAFT CMs, as well as the SR system. In Section IV, we show the design details of S-PBFT and S-RAFT. We derive and analyze the consensus success rate, energy consumption, latency, and communication overhead of these two CMs in Section V. Then, in Section VI we simulate the above performance of 6 PBFT-like and 4 RAFT-like CMs modified by SBC. Finally, Section VII concludes this work.

## II. RELATED WORK

In this section, PBFT-like CM, RAFT-like consensus, and wireless blockchain consensus are reviewed respectively.

### A. PBFT-like CM

PBFT is a typical Byzantine fault-tolerant (BFT) consensus that can run a network with up to 1/3 of Byzantine nodes [10]. Due to its high Byzantine fault tolerance, it is considered to be one of the most potential applications of BFT consensus.

However, since the complicated consensus process of PBFT, especially the two broadcast processes, it has $O(n^2)$ communication overhead [25], which limits the latency, throughput, and scalability. To solve these defects and promote its deployment in various network scenarios, many researchers have proposed improvement schemes for PBFT.

We refer to these schemes as PBFT-like CMs.

To improve the consensus efficiency, Gao *et al.* [26] have designed Trust PBFT (T-PBFT), which removes the second broadcast from PBFT and selects multiple nodes as the primary node group based on EigenTrust to replace a single node in the PBFT. Moreover, Xu *et al.* [27] have designed Score Grouping-PBFT (SG-PBFT). It divides the nodes into consensus nodes and candidate nodes by the grouping score mechanism, and the consensus process is carried out only in the consensus nodes. This CM also reduces a broadcast process to improve consensus efficiency. On this basis, in [28], the authors have used an artificial bee colony (ABC) algorithm to make group scoring more efficient, and proposed ABC-PBFT. Meanwhile, in [29], the authors have optimized the broadcast process in PBFT using a rollback mechanism and proposed Votes-as-a-Proof (VaaP).

In addition, there are many schemes to improve the PBFT scalability, among which the typical representative is the hierarchical blockchain, such as Multilayer PBFT [25], bottom-to-up hierarchical chain (B2UHChain) [30], ultra-low storage overhead PBFT (ULS-PBFT) [31], diffusion PBFT (DPBFT) [32], NBFT [33], and so on. These methods are similar to sharding. They group nodes into regular groups, and then select committee nodes from each group to achieve global consistency. They can limit communication overhead, storage overhead, etc. within each group to improve consensus scalability. In particular, the DPBFT is a committee consensus first before starting with the consensus in each group. In contrast, NBFT first conducts consensus within each group and finally reaches consistency in the committee.

*B. RAFT-like CM*

RAFT is a typical Crash fault-tolerant (CFT) consensus that can run a network with up to 1/2 failed nodes [11]. Its implementation method is simpler than the original CFT consensus Paxos [34]. Thus, it is widely used in many private chain scenarios.

However, the implementation of RAFT often relies on the normal work of the leader, who assumes the responsibility of publishing consensus messages to all followers, with a large communication load. In addition, when the leader fails, it is also necessary to restart the leader election, which wastes much time. To solve this problem, many researchers have proposed optimization schemes. We call such schemes RAFT-like CMs.

Wang *et al.* [35] have designed Kademlia-RAFT (KRAFT) where multiple leaders can share the communication load of delivering consensus messages to followers. Multiple leaders can also avoid the single-point failure of a single leader. Additionally, in [20], the authors have proposed that it is possible to set part of the followers as two-hop followers based on the node geographical location, namely, two-hop RAFT (TH-RAFT). Then, the communication of these two-hop followers will be completed by other single-hop followers, thus reducing the communication load of the leader. Meanwhile, Tian *et al.* [36] have designed verifiable secret sharing Byzantine fault tolerance-RAFT (VSSB-RAFT) by combining zero-knowledge proof with RAFT to make RAFT also have Byzantine fault-tolerance capability. They also use the named-data-networking forwarding daemon (NFD) to divide the 10 consensus nodes into a communication domain. Each domain is responsible for releasing consensus messages for its followers, reducing the burden on the leader.

Moreover, some other work focuses on the leader election process at RAFT to elect a leader more efficiently, such as the schemes from [37-38]. However, these schemes do not involve the communication process in RAFT, and thus they are not the main focus and improvement of our SBC.

*C. Wireless Blockchain Consensus*

Wireless blockchain consensus is often applied to WBNs. This concept comes from when the blockchain is deployed in wireless networks or when the blockchain is networked by wireless communications. However, blockchain CMs were originally designed for wired networks [17-18]. Therefore, traditional CMs are difficult to apply to WBNs, and it is urgent to design a dedicated wireless blockchain consensus.

Currently, one aspect of the work focuses on the performance of various CMs in wireless networks. For example, Zhang *et al.* [39] have studied the performance of wireless PBFT, RAFT, and Proof of Work (PoW) consensus, including communication complexity, transmission power, consensus success rate. Then, in [40], the authors researched the consensus range of the wireless RAFT consensus in the presence of malicious nodes. Moreover, some works focus on the performance of blockchain CMs in different wireless scenarios, such as IoV [4] and [41], 6G wireless networks [15-16], unmanned aerial vehicle networks [42], etc.

Additionally, another aspect of the work is dedicated to improving the performance of wireless blockchain consensus. Xu *et al.* [17-18] have designed efficient and reliable consensus protocols for single-hop and multi-hop wireless networks, respectively, taking into account the actual Signal-to-Interference-plus-Noise-Ratio (SINR) model. Based on this, Zou *et al.* [43] have considered the existence of Byzantine nodes in wireless networks and designed a fast CM. In [44], the authors have proposed a CM to resist Sybil attacks in wireless networks.

However, the above work lacks the consideration and utilization of wireless network spectrum resources. Spectrum resources are the key to support wireless communications. We plan to deploy SR in the WBN, make full use of spectrum resources, and design the SBC to improve the applicability of blockchain CMs in wireless networks.

III. PRELIMINARIES

In this section, we introduce PBFT and RAFT consensus, and the SR system in turn. The purpose of introducing PBFT and RAFT is to demonstrate the SBC design scheme with these two CMs as representatives. The frequently used notations are listed in Table I.

*A. PBFT and RAFT Consensus*

*1) PBFT Consensus*

The consensus process of PBFT is shown in Fig. 1. The roles that participate in the consensus process include client, primary node, and replica. After the client sends a consensus request to the primary node, the PBFT will start the four phases, including *pre-prepare*, *prepare*, *commit*, and *reply*.

TABLE I
FREQUENTLY USED NOTATIONS

| Notation | Definition |
| --- | --- |
| $n$ | Total number of nodes in the WBN |
| $b$ | Number of Byzantine nodes in a wireless PBFT CM |
| $f$ | Number of failed nodes in a wireless PBFT CM |
| $P_s$ | Transmission success rate not enhanced by backscattering |
| $P_e$ | Transmission success rate of enhanced communications |
| $P_{S\text{-}PBFT}$ | Consensus success rate of S-PBFT |
| $P_{S\text{-}RAFT}$ | Consensus success rate of S-RAFT |
| $L(FD)_{S\text{-}PBFT}$ | Consensus latency of S-PBFT with FD multiplexing |
| $L(TD)_{S\text{-}PBFT}$ | Consensus latency of S-PBFT with TD multiplexing |
| $L(FD)_{S\text{-}RAFT}$ | Consensus latency of S-RAFT with FD multiplexing |
| $L(TD)_{S\text{-}RAFT}$ | Consensus latency of S-RAFT with TD multiplexing |
| $C_{S\text{-}PBFT}$ | Communication overhead of S-PBFT |
| $C_{S\text{-}RAFT}$ | Communication overhead of S-RAFT |
| $E_{S\text{-}PBFT}$ | Energy consumption of S-PBFT |
| $E_{S\text{-}RAFT}$ | Energy consumption of S-RAFT |

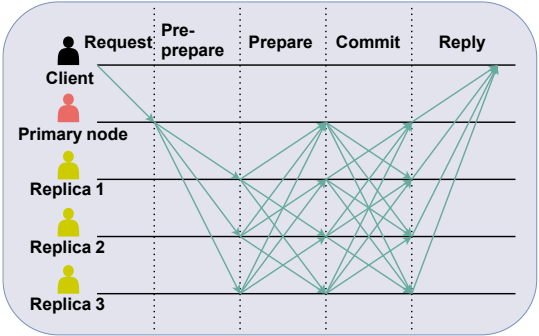

**Fig. 1.** PBFT consensus.

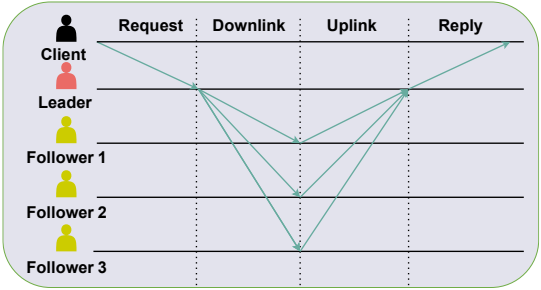

**Fig. 2.** RAFT consensus.

The *prepare* and *commit* phases require multiple nodes to broadcast consensus messages and have many communication processes [1]. Therefore, these two phases are also the parts of PBFT that incur a major communication overhead, latency, and energy consumption. According to the conclusion in [16], the consensus energy consumption of PBFT presents a cubic increase concerning the number of nodes.

For the fault tolerance of PBFT, it can accommodate up to 1/3 Byzantine or failed nodes. If the total number of nodes in the WBN is $n$ and the number of Byzantine or faulty nodes is $b$, then the following conditions must be satisfied to ensure a successful consensus:

$$b \leq \left\lfloor \frac{n-1}{3} \right\rfloor. \qquad (1)$$

*2) RAFT Consensus*

The specific consensus process of RAFT is shown in Fig. 2. The roles that participate in the consensus process include client, leader, and follower (there will be candidates during the leadership election) [4], [14], [16], [40]. After the client sends a consensus request to the leader, the leader sends consensus messages to all followers by *downlink*. The followers then vote and report their results back to the leader by *uplink*. Finally, the leader returns a reply message to the client.

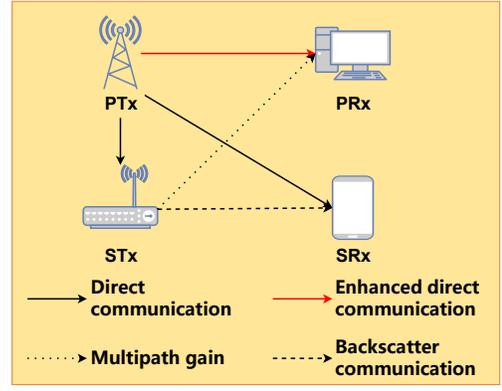

**Fig. 3.** Symbiotic radio system.

It is clear that the two main phases in RAFT that affect consensus performance are *downlink* and *uplink* transmission [4], [16]. Although RAFT has a lower consensus energy consumption than PBFT, it still shows a square increase in the number of nodes [16].

For the fault tolerance of RAFT, it can accommodate up to 1/2 failed nodes. If the total number of nodes in the WBN is $n$ and the number of faulty nodes is $f$, then the following conditions must be satisfied to ensure a successful consensus:

$$f \leq \left\lfloor \frac{n-1}{2} \right\rfloor. \qquad (2)$$

*B. Symbiotic Radio System*

The birth of SR provides technical support for WBN to make full use of its spectrum and energy resources and improve consensus performance. The realization of SR based on cognitive backscatter communication technology is inseparable from the antenna scattering principle and modulation in the air, which are not the main focuses of SBC and will not be covered here. We can see [23-24], [45] for details. According to the SR system in [45], it consists of a primary system and a secondary system, as shown in Fig. 3. Further, the primary system includes a primary transmitter (PTx), and a primary receiver (PRx). The secondary system has a secondary transmitter (STx), and a secondary receiver (SRx).

This system is expected to achieve a mutualistic relationship and symbiotic paradigm. Specifically, in the primary system, the PTx sends information to the PRx via active transmission. In the secondary system, the STx uses the RF signal emitted from the primary system to send information to the SRx by passive backscatter transmission. In this way, the signals received by the PRx include direct communications from the PTx and backscatter communications from the STx. Moreover, the signal in the backscatter communication contains the PTx transmission information. Therefore, the PRx is expected to achieve the multipath gain provided by the secondary system and improve its communication performance (i.e., enhanced direct communication). Meanwhile, the STx transmits information to the PRx using the RF signal from the PTx, which is a passive transmission and can reduce energy consumption [23-24], [46]. Based on the above advantages, SR is listed as one of the key candidate technologies for 6G, along with blockchain [47].

Additionally, the symbol from the secondary system is multiplied by the primary system sent symbol. Thus, the symbol period ratio of the secondary system to the primary system is called the spreading factor $K$. Also, the $K$ should meet a certain condition to construct the symbiotic system [23-24], namely

$$K \geq \frac{\left(f_Q^{-1}\left(\frac{f_Q\left(\sqrt{M\gamma_d}\right) - f_Q\left(\sqrt{M\gamma_d(1+\Delta\gamma)}\right)}{f_Q\left(\sqrt{\frac{M\gamma_d}{1+\Delta\gamma}}(1-\Delta\gamma)\right) - f_Q\left(\sqrt{M\gamma_d(1+\Delta\gamma)}\right)}\right)\right)^2}{\varpi^2}, \quad (3)$$

$$\varpi = \frac{\sqrt{2M\Delta\gamma}\left(1 - 2f_Q\left(\sqrt{M\gamma_d}\right)\right)}{\sqrt{1/\gamma_d + 4M\Delta\gamma f_Q\left(\sqrt{M\gamma_d}\right)\left(1 - f_Q\left(\sqrt{M\gamma_d}\right)\right)}}, \quad (4)$$

where $f_Q$ is the $Q$ function, and $\gamma_d$ represents the average signal-to-noise ratio (SNR) of the direct communication. The $\Delta\gamma = \gamma_b/\gamma_d$ denotes the relative SNR between the direct communication and the backscatter communication. In particular, $\gamma_b$ is the average SNR of the backscatter communication. $M$ represents the number of antennas, and $\varpi$ denotes a variable about $\gamma_d$, $\Delta\gamma$, and $M$.

Furthermore, [23-24] have verified the above (3), and (4) through simulations, and give the symbiotic law of primary and secondary systems (P&S), which will be used in SBC: 1) When $\gamma_d$ and $K$ are large enough, the SNR of primary system $\gamma_P$ is the sum of $\gamma_d$ and $\gamma_b$; 2) When $\gamma_d$ is large enough, the secondary system of SNR $\gamma_S$ is $K$ times $\gamma_b$. These two can be expressed by (5) and (6). More details of the above can refer to [23-24].

$$\gamma_P = \gamma_d + \gamma_b, \quad (5)$$
$$\gamma_S = K\gamma_b. \quad (6)$$

## IV. MODEL DESIGN

In this section, we introduce SBC and use S-PBFT and S-RAFT as examples to demonstrate the specific design methods of SBC. It covers the necessary changes in the role of SR in SBC and the consensus process for S-PBFT and S-RAFT.

### A. Symbiotic Blockchain Consensus

SBC is essentially the deployment of SR systems in wireless blockchain consensus to achieve mutualistic transmission between consensus nodes in the WBN. This transmission relationship will use passive communication[2] to replace part of the active communication to achieve energy-saving effects, while the multipath gain will also enhance the signal strength of the active communication, thus achieving a high consensus success rate and low energy consumption of the WBN. In this part, we will explain the necessary improvements required when implementing SBC in WBN.

---

[2] Passive communication refers to backscatter communication in this paper.

First, the devices in the SR and nodes in the WBN have different functions, and thus we need to make necessary modifications to the device in the SR to realize the symbiotic SBC. For the SR proposed in [22-24], the primary system is an active communication device, while the secondary system is a passive communication device. However, the passive device cannot actively transmit data, which clearly does not meet the needs of nodes in blockchain networks, because these nodes will actively communicate with each other. In the WBN, each node is set as an active communication device with active communication capability, and a passive scattering antenna is installed on each node to realize backscatter communications. In other words, each node in the WBN has two types of antennas, namely active antenna and passive antenna. Therefore, every node can be used as a primary system device and a secondary system device in the SR. This setup means every node can also use the low-energy-consumption antenna impedance and load impedance to enhance the passive signal strength [45]. It should be noted that passive antennas are very low-cost and do not impose a cost burden on SBC.

Second, as a CM deployed in a wireless network, the SBC has multiple pairs of P&S communicating simultaneously. Therefore, communication interference may be generated between these P&S, affecting the communication performance and reducing the consensus reliability of SBNs. To avoid this problem, we use frequency division (FD) or time division (TD) multiplexing according to spectrum resources, so that each pair of P&S uses a specific frequency band (FB) or time slot. Different P&S use different frequency bands or time slots. In this way, the SR system does not interfere with each other, thus, jointly implementing the SBC.

### B. S-PBFT Consensus

PBFT-like CMs are almost always modified and optimized based on PBFT. Therefore, we take PBFT as a representative to illustrate the design ideas and details of S-PBFT.

First, when spectrum resources are sufficient, we can assign a separate FB to each SR system. Then, the S-PBFT consensus with FD multiplexing to achieve symbiosis characteristics is shown in Fig. 4 (taking $n$=4 as an example). Specifically, it can be described by the following steps:

- *Request:* The client sends a consensus *request* to the primary node.
- *Pre-prepare:* In this phase, the primary node, acting as PTx for Replicas 1, 2, and 3, sends a *pre-prepare* message to these three replicas. Then, these three replicas have both STx and PRx roles, where PRx is the correspondence between these replicas and PTx, and STx is the relative relationship between these replicas. Specifically, as Replica 2 of STx transmits multipath gain to Replica 1 of PRx via backscatter communication; as Replica 3 of STx transmits multipath gain to Replica 2 of PRx via backscatter communication; as Replica 1 of STx transmits multipath gain to Replica 3 of PRx via backscatter communication. As a result, the *pre-prepare* messages sent by the primary node (PTx) to all three replicas are all enhanced.
- *Prepare:* There is a continuation of backscatter communication from the previous phase. In this case, replicas and the primary node play the SRx and STx with each other, namely, as Replica 1 of STx transmits the *prepare* message to Replica 2 of SRx by modulating the RF signal from PTx; as Replica 2 of STx transmits the *prepare* message to Replica 3 of SRx by modulating the RF signal from PTx; as Replica 3 of STx transmits the *prepare* message to Replica 1 of SRx by modulating the RF signal from PTx. The non-energy-consuming backscatter communications can replace the *prepare* message transmissions in this phase, thus achieving the energy-saving effect. Backscatter communications can replace $n$-1 *prepare* messages. Additionally, each replica will actively communicate the remaining $n^2$-$3n$+2 *prepare* messages. The generated active communications further provide multipath gain for other replicas, which is similar to the *pre-prepare*.

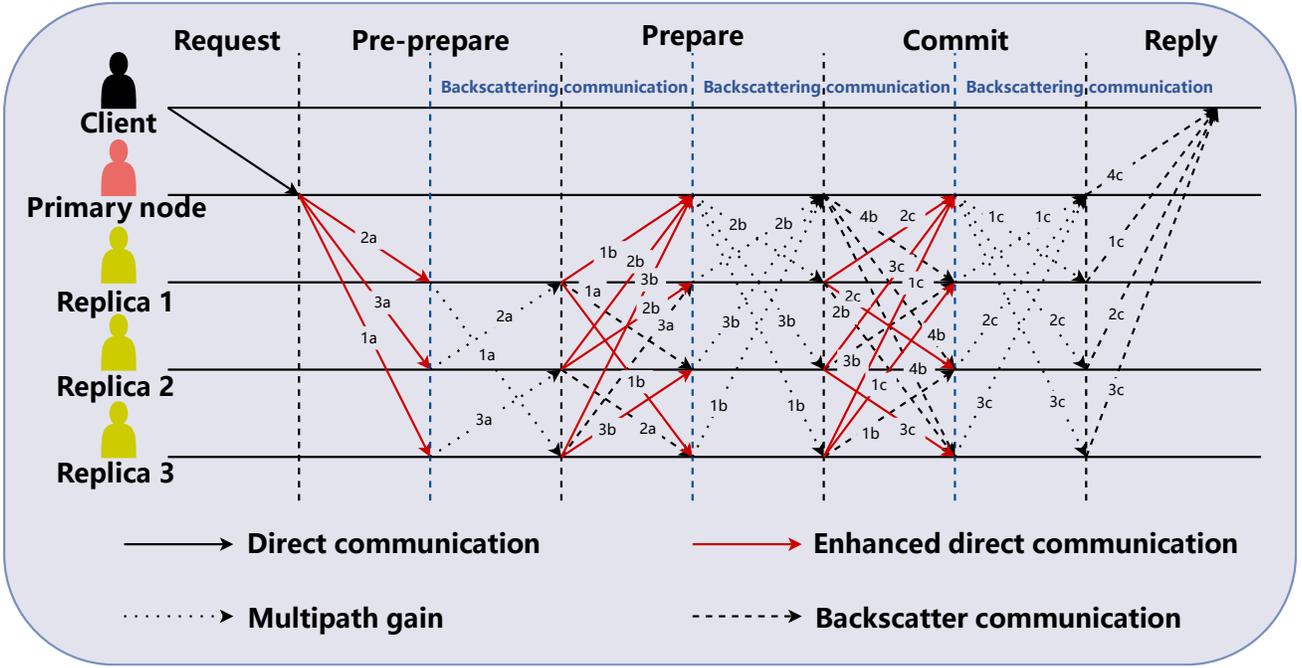

**Fig. 4.** S-PBFT consensus with FD multiplexing.

- *Commit:* Similar to the previous phase, there is also a continuation of backscatter communications, which can be substituted for the $2n$-2 *commit* messages. We modulate these scattered signals to a new FB to avoid interference with active communications at this phase. Then, each replica will actively communicate the remaining $n^2$-$3n$+2 *commit* messages. These active communications also cause the passive antennas of each STx to contribute multipath gains to the other replicas, enhancing the reliability of active communications.
- *Reply:* The *reply* messages at this phase can be completely replaced by backscatter communications, so this phase is energy-free.

It should be noted that in the consensus process, some phase nodes accept multiple backscatter signals or multipath gains from different STx simultaneously. This model is called multiuser multi-backscatter-device SR (MuMB-SR), and the receiver set can be referred to [48].

Moreover, in Fig. 4, the numbers "1, 2, 3, 4" represent the FBs used by different P&S, avoiding the same frequency interference. The letters "a, b, c" indicate the time sequence of communications. As a result, the "1a" is the communication using FB "1" when the time slot is "a". For the *prepare* and *commit* phases, there are two broadcast processes. Thus, we set the PTx using one FB broadcast to save bandwidth resources. In our statistics, S-PBFT uses $n$ FBs at most. Then, taking the *pre-prepare* phase as an example, we use Fig. 5 to reveal the symbiotic relationship between these nodes, and Table II shows the roles played by each node in different FBs.

TABLE II
THE ROLE OF EACH NODE IN THE *PRE-PREPARE* PHASE

| FBs | Primary node | Replica 1 | Replica 2 | Replica 3 |
|---|---|---|---|---|
| 1 | PTx | STx | SRx | PRx |
| 2 | PTx | PRx | STx | SRx |
| 3 | PTx | SRx | PRx | STx |

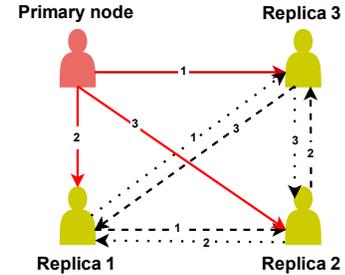

**Fig. 5.** Symbiotic relationship between each node in the *pre-prepare* phase.

Second, when the spectrum resources are insufficient, we have to use TD multiplexing technology to avoid the same frequency interference between different SR systems. It is an unfortunate tradeoff between time and spectrum. In particular, the broadcasts initiated by different nodes in the *prepare* and *commit* phases will take place in different time slots. Since space constraints, a diagram of S-PBFT with TD multiplexing similar to Fig. 4 is not provided here. Clearly, we think S-PBFT will have a higher consensus latency under TD multiplexing than under FD multiplexing due to the multiple broadcasts. This will be analyzed in Section V and demonstrated in Section VI.

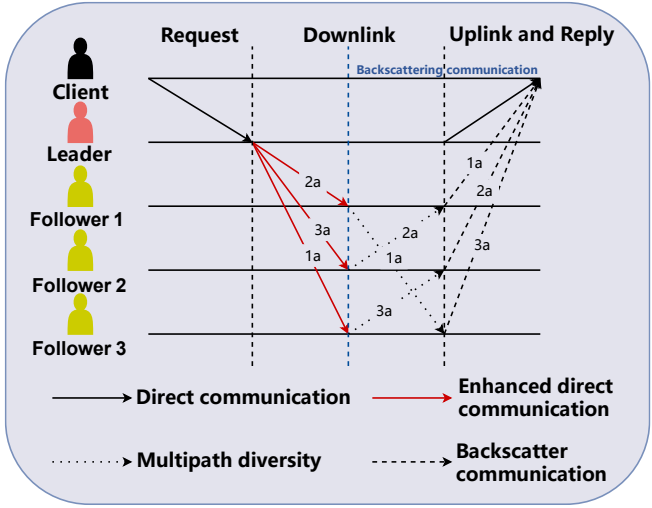

**Fig. 6.** S-RAFT consensus with FD multiplexing.

### C. S-RAFT Consensus

Similar to S-PBFT, we take RAFT as a representative to illustrate the design ideas and details of S-RAFT.

First, we also consider assigning a separate FB to each SR system in S-RAFT when spectrum resources are sufficient. Then, the S-RAFT consensus process with FD multiplexing is shown in Fig. 6 (taking $n$=4 as an example). Compared with the original RAFT, S-RAFT combines uplink and reply, benefiting from the backscatter properties of SR. In this way, the consensus process can be simplified. Like PBFT, the final consensus result judgment is left to the client. The normal operation of PBFT provides a basis for our setting. Specifically, S-RAFT can be described by the following steps:

- *Request:* The client sends a consensus *request* to the Leader.
- *Downlink:* This phase is similar to *pre-prepare* in S-PBFT. The Leader acting as PTx for Followers 1, 2, and 3, sends consensus messages to these three by *downlink* communications. The three replicas have both STx and PRx roles, where PRx is the correspondence between these followers and PTx, and STx is the relative relationship between these followers. Specifically, as Follower 2 of STx transmits multipath gain to Follower 1 of PRx via backscatter communication; as Follower 3 of STx transmits multipath gain to Follower 2 of PRx via backscatter communication; as Follower 1 of STx transmits multipath gain to Follower 3 of PRx via backscatter communication. Consequently, the consensus messages sent by the Leader (PTx) to three followers are enhanced.
- *Uplink* and *Reply:* This phase is similar to the *reply* in S-PBFT. In particular, in addition to the Leader replying to the client through *uplink* communication, the feedback of the three followers to the consensus message is made by backscatter communication. At this time, the followers play STx and the client plays SRx.

Consequently, compared with RAFT, S-RAFT can save $n$ communication overhead. In particular, the *uplink* communication part is replaced by backscatter communication. In S-RAFT, the roles of each node are similar to those in Fig. 5 and Table II of S-PBFT, except that the primary node (S-PBFT) becomes the leader (S-RAFT) and replicas (S-PBFT) become followers (S-RAFT).

Second, TD multiplexing technology must be adopted when S-RAFT has insufficient spectrum resources to avoid the same frequency interference between different SR systems. That is, different SR systems in S-RAFT need to communicate in different time slots. Similar to S-PBFT with TD multiplexing, due to space constraints, this paper does not provide a diagram of S-PBFT with TD multiplexing similar to Fig. 6. The performance of this part will be also explained in Sections V and VI.

## V. PERFORMANCE ANALYSIS

To quantitatively characterize the consensus performance of SBC, we provide a comprehensive performance analysis method for S-PBFT and S-RAFT, including consensus success rate, consensus energy consumption, consensus latency, and communication overhead.

### A. Performance Analysis of S-PBFT Consensus

#### 1) Consensus Success Rate

Compared with the PBFT, the S-PBFT has enhanced active communications and passive backscatter communications. In [51], the authors provide a scheme to minimize the bit error rate of the secondary system, so we consider that the backscatter communication reliability is in line with PBFT [1], and then we set its transmission success rate of a single channel as $P_s$. We further set the transmission success rate of enhanced active communications in S-PBFT to be $P_e$.

Inspired by [15-16], we can find the consensus success rate of PBFT in wireless networks. Then, for the *pre-prepare* phase of S-PBFT, the success rate is

$$P_1 = \sum_{i=0}^{b} C_{n-1}^{i} (1-P_e)^i P_e^{(n-1-i)}. \tag{7}$$

For the *prepare* phase, there exists $n$-1 enhanced active communications for the primary node as a receiver. Thus, the success rate is

$$P_2 = \sum_{j=0}^{b-i} C_{n-1-i}^{j} (1-P_e)^j P_e^{(n-1-i-j)}. \tag{8}$$

For the *commit* phase, there also exist $n$-1 enhanced active communications for the primary node as a receiver. Thus, the success rate is

$$P_{PN} = P_2. \tag{9}$$

Additionally, there exist $n$-3 enhanced active communications and 2 passive communications for the $n$-1 replicas. Thus, the success rate for them is

$$P_{Re} = \sum_{k=0}^{b-2} C_{n-3}^{k} (1-P_e)^k P_e^{(n-3-k)} \\ + C_{n-3}^{b-1}(1-P_e)^{b-1} P_e^{(n-b-2)} C_2^1 P_s (1-P_s) \tag{10} \\ + C_{n-3}^{b-2}(1-P_e)^{b-2} P_e^{(n-b-3)} P_s^2.$$

As a result, the success rate for this phase is

$$P_3 = \sum_{l=0}^{b-1} C_{n-1}^{l} (1-P_{Re})^l P_{Re}^{(n-1-l)} \\ + C_{n-1}^{b-1}(1-P_{Re})^{b-1} P_{Re}^{(n-b)} P_{PN}. \tag{11}$$

$$P_{S-PBFT} = \sum_{i=0}^{b} \left( C_{n-1}^{i} (1-P_e)^{i} P_e^{(n-1-i)} \sum_{j=0}^{b-i} \left( C_{n-1-i}^{j} (1-P_e)^{j} P_e^{(n-1-i-j)} \right) \right)$$
$$\sum_{l=0}^{b} \left( \left( C_{n-1}^{l} (1-P_{Re})^{l} P_{Re}^{(n-1-l)} + C_{n-1}^{b+1} (1-P_{Re})^{b+1} P_{Re}^{(n-b-2)} P_{PN} \right) \sum_{m=0}^{b-l-m} C_{n-l}^{m} (1-P_e)^{m} P_e^{(n-l-m)} \right). \quad (13)$$

For the *reply* phase, there exists $n$ enhanced active communications. Thus, the success rate is

$$P_4 = \sum_{m=0}^{b-l-m} C_{n-l}^{m} (1-P_e)^{m} P_e^{(n-l-m)}. \quad (12)$$

In general, the consensus success rate of wireless S-PBFT is closely related to (7), (8), (11), and (12), and can be expressed as (13) at the up of this page.

The derive of (10) and (11) are given in Appendix.

*2) Consensus Latency*

The premise of analyzing consensus energy consumption is to clarify the consensus latency. Thus, we analyze its latency in advance.

First, for the FD multiplexing, according to [16], the latency of PBFT is

$$L(FD)_{PBFT} = 3t_1 + t_2, \quad (14)$$

where $t_1$ represents the latency of each first three phases, and $t_2$ denotes the latency of the *reply* phase. In other words, the latency of the first three phases of PBFT is the same.

Inspired by [15-16], [49-50] the values of $t_1$ and $t_2$ can be obtained as follows:

$$1 - P_s = f_Q \left( \frac{NTBC - NTBR + \frac{\log NT_s B}{2}}{(\log e)\sqrt{NTB}} \right), \quad (15)$$

where $T_s$ is the latency for a no-enhanced signal, which is related to the number of channels connected by the transmission node, namely $(n-1)T_s = t_1$ and $T_s = t_2$. $B$ represents the bandwidth. $R$ and $C$ are the transmission rate and channel capacity, respectively. $N$ represents the number of subcarriers, in our paper $N=1$.

For S-PBFT, the signals of the first three phases are enhanced, so the latency of these phases can be expressed as

$$1 - P_e = f_Q \left( \frac{NTBC_e - NTBR + \frac{\log NT_e B}{2}}{(\log e)\sqrt{NTB}} \right), \quad (16)$$

where $C_e$ represents the channel capacity enhanced by multipath gain, and $T_e$ denotes latency for an enhanced signal. Thus, the latency for each first three phases of S-PBFT $t_3$ is $(n-1)T_e$. Similarly, the *reply* phase latency $t_4$ of S-PBFT equals $T_e$. As a result, we can see that the S-PBFT consensus latency is

$$L(FD)_{S-PBFT} = 3t_3 + t_4. \quad (17)$$

Second, for the TD multiplexing, the broadcast processes on each node cannot perform simultaneously in the *prepare* and *commit* phases. Therefore, for PBFT, the latency of these two phases is $(n-1)t_1$ and $nt_1$ respectively; For S-PBFT, the latency of two phases is $(n-1)t_3$ and $nt_3$ respectively. Therefore, the consensus latency of PBFT and S-PBFT can be obtained by

$$L(TD)_{PBFT} = 2nt_1 + t_2, \quad (18)$$
$$L(TD)_{S-PBFT} = 2nt_3 + t_4. \quad (19)$$

*3) Communication Overhead*

Communication overhead refers to the number of communications required for consensus, also known as communication complexity or cost [4], [25], [27]. For S-PBFT, only active communication is included in the statistics of communication overhead, because passive communication does not consume communication bandwidth and energy.

For the four phases of S-PBFT, it has $n-1$, $n^2-3n+2$, $n^2-3n+2$, and 0 active communications respectively. Therefore, its communication overhead can be expressed as

$$C_{S-PBFT} = 2n^2 - 5n + 3. \quad (20)$$

*4) Energy Consumption*

According to [16], we know the energy consumption of the PBFT consensus is

$$E_{PBFT} = (2n^2 t_1 - 2nt_1 + nt_2) P_T, \quad (21)$$

where $P_T$ is the transmitting power.

In this version, to measure the energy consumption more accurately, we consider the positive effect of multipath gain on consensus latency.

As a consequence, for S-PBFT, the energy consumption of the *pre-prepare* stage is

$$E_1 = (n-1)t_3 P_T. \quad (22)$$

In the *prepare* stage, there exist $n^2-3n+2$ enhanced active communications. So, the energy consumption of this stage is

$$E_2 = (n^2 - 3n + 2)t_3 P_T. \quad (23)$$

The *commit* stage is the same as the *prepare* stage. Thus, the energy consumption $E_3$ equals $E_2$.

At last, for the *reply* stage, there are all energy-free passive backscatter communications without energy consumption.

Therefore, the energy consumption of S-PBFT consensus is

$$E_{S-PBFT} = E_1 + E_2 + E_3 = (2n^2 - 5n + 3)t_3 P_T. \quad (24)$$

B. Performance Analysis of S-RAFT Consensus

*1) Consensus Success Rate*

Compared to RAFT, S-RAFT enhances the *downlink* phase of consensus, and similar to S-PBFT, the backscatter communication reliability at the *uplink* phase can also be consistent with RAFT. Then, the consensus success rate of S-RATF can be expressed as the product of the success rates of these two communication phases [16].

For the *downlink* communication, based on a 50% fault tolerance rate [11], the success rate of this phase is

$$P_{downlink} = \sum_{i=0}^{f} C_{n-1}^{i} (1-P_s)^{i} P_s^{(n-1-i)}. \quad (25)$$

To ensure a successful S-RAFT consensus, the maximum communication failure allowed for *uplink* communication is $f-i$,

since there have already been $i$ failures in the *downlink* communication. Then, the success rate of this phase is

$$P_{uplink} = \sum_{j=0}^{f-i} C_{n-1-i}^{j} \left(1-P_e\right)^j P_e^{(n-1-i-j)}. \tag{26}$$

In general, the consensus success rate of wireless S-RAFT is closely related to (25), and (26), and can be expressed as (27).

$$P_{S-RAFT} = \sum_{i=0}^{f} \left( C_{n-1}^{i} \left(1-P_s\right)^i P_s^{(n-1-i)} \sum_{j=0}^{f-i} C_{n-1-i}^{j} \left(1-P_e\right)^j P_e^{(n-1-i-j)} \right). \tag{27}$$

*2) Consensus Latency*

As with S-PBFT, we next analyze the consensus latency and throughput of S-RAFT.

First, for the FD multiplexing, according to [16], the latency of RAFT is

$$L(FD)_{RAFT} = t_1 + t_2. \tag{28}$$

Therefore, the consensus latency of S-RAFT is

$$L(FD)_{S-RAFT} = t_3 + t_4. \tag{29}$$

Second, since there are no multiple parallel broadcast processes in S-RAFT, the consensus latency of TD multiplexing is the same as that of FD multiplexing, namely $L(FD)_{S-RAFT}=L(TD)_{S-RAFT}$.

*3) Communication Overhead*

As with S-PBFT, the communication overhead of S-RAFT also involves only active communications. Due to the *uplink* phase is passive communication, its communication cost is

$$C_{S-RAFT} = n. \tag{30}$$

*4) Energy Consumption*

Since S-RAFT only has downlink phase active communication, its total energy consumption is

$$E_{S-RAFT} = n t_3 P_T. \tag{31}$$

## VI. PERFORMANCE EVALUATION

The previous section is the performance analysis for S-PBFT and S-RAFT, which can also be generalized to PBFT-like and RAFT-like CMs. To highlight the universality of SBC, we select 10 SOTA CMs and evaluate their performance due to their focus on optimizing the consensus communication process. Table III provides their brief description. These CMs are S-PBFT, Symbiotic T-PBFT (ST-PBFT), Symbiotic ABC-PBFT (SABC-PBFT), Symbiotic VaaP (S-VaaP), Symbiotic DPBFT (S-DPBFT), Symbiotic NBFT (S-NBFT), S-RAFT, Symbiotic TH-RAFT (STH-RAFT), Symbiotic KRAFT (S-KRAFT), and Symbiotic VSSB-RAFT (SVSSB-RAFT) in turn. Since the unique design of some CMs, we set the proportion of primary nodes in T-PBFT to 20%; there are 5 shards in both DPBFT and NBFT; the proportion of consensus nodes in ABC-PBFT is 40%; KRAFT has five leaders; TH-RAFT has 20% two-hop nodes. Meanwhile, according to settings from [15-16], [23-24], [52], we assume that the bandwidth of each FB is 1 MHz, and $R$=100 kbps, $C$=150 kbps, $P_T$=30 dBm. The simulation is running with MATLAB R2021a on a PC equipped with an Intel I7-1260P processor, which has a clock frequency of 2.1 GHz and 16 GB RAM.

TABLE III
CONSENSUS MECHANISMS IMPROVED BY SBC

| CM type | CM name | Brief introduction |
| --- | --- | --- |
| PBFT-like | PBFT [10] | The original PBFT consensus |
| | T-PBFT [26] | Replace a single primary node with a primary node group |
| | ABC-PBFT [28] | Distinguish the consensus nodes from all nodes |
| | VaaP [29] | Improve parallel broadcasting with a rollback mechanism |
| | DPBFT [32] | Multi-group parallel consensus starting from the committee |
| | NBFT [33] | Multi-group parallel consensus starting from each shard |
| RAFT-like | RAFT [11] | The original RAFT consensus |
| | TH-RAFT [20] | Two-hop RAFT consensus |
| | KRAFT [35] | Replace a single leader with multiple leaders |
| | VSSB-RAFT [36] | The blockchain network is divided into groups of 10 nodes |

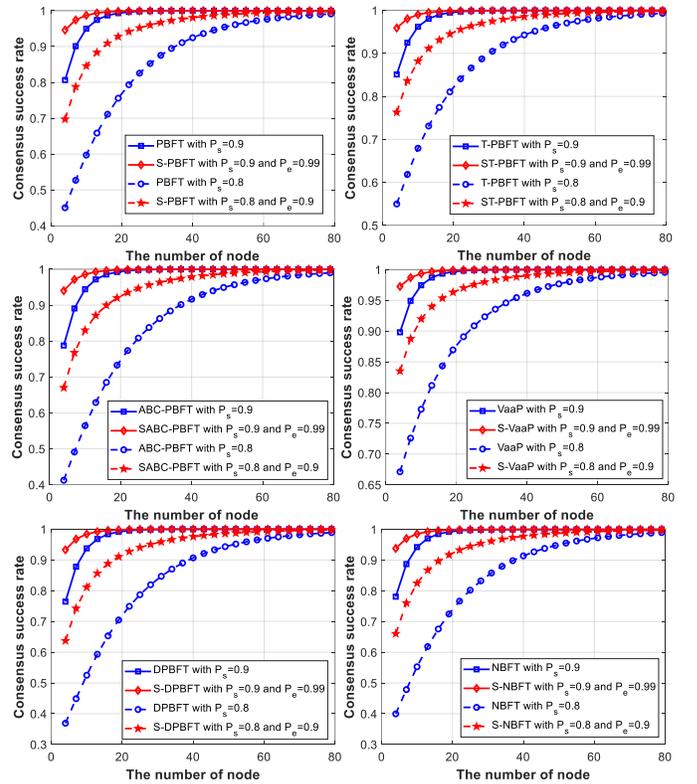

**Fig. 7.** Consensus success rate of PBFT-like CMs.

*A. Performance Evaluation of PBFT-like CM*

First, we have evaluated the **consensus success rate** of PBFT-like CMs improved by multipath gain. We set $P_s$=0.9, $P_e$= 0,99; and $P_s$=0.8, $P_e$=0.9 to show the gain strength of SR for different transmission success rates shown in Fig. 7. Both results indicate that all SBCs benefited from the backscatter communication, and have a higher consensus success rate than the original CMs. The enhancement of the consensus success rate is most prominent when the number of nodes is small because the consensus is more reliable when the number of nodes is large. When $n$=4, $P_s$=0.8, and $P_e$=0.9, the consensus success rate of all 6 PBFT-like SBCs increases by an average of 54.1%. These simulation results can also show the performance and differences of different PBFT-like in wireless networks, including the original design and the design improved by SR. Among them, ABC-PBFT has a lower consensus success rate than the original PBFT, because it only selects part nodes in the whole network to participate in consensus. Thanks to a more trusted and simplified process, T-PBFT and VaaP have achieved a higher consensus success rate. In addition, DPBFT and NBFT demonstrate the interesting fact that CM starting with each shard has a higher consensus success rate than starting with a committee. This result is consistent with the conclusion in B2UHChain [30].

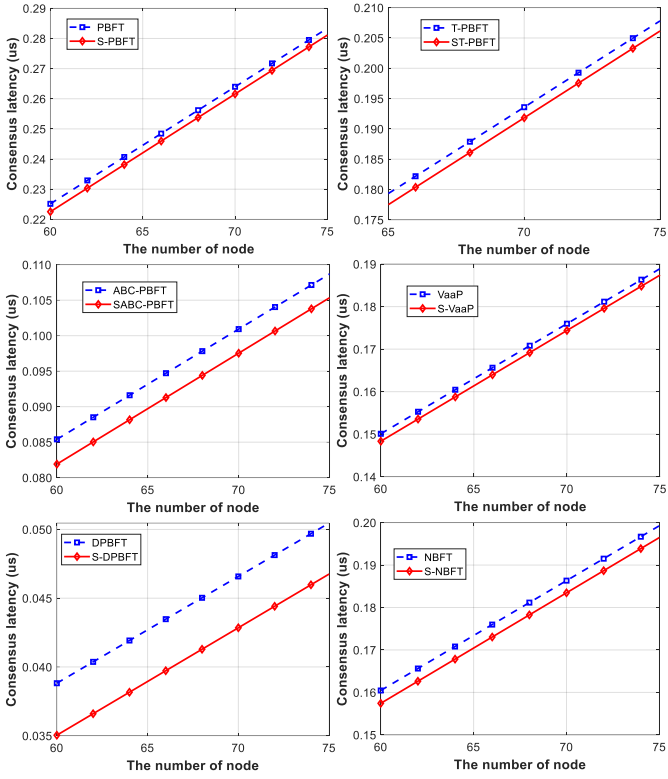

**Fig. 8.** Consensus latency of PBFT-like CMs under FD multiplexing.

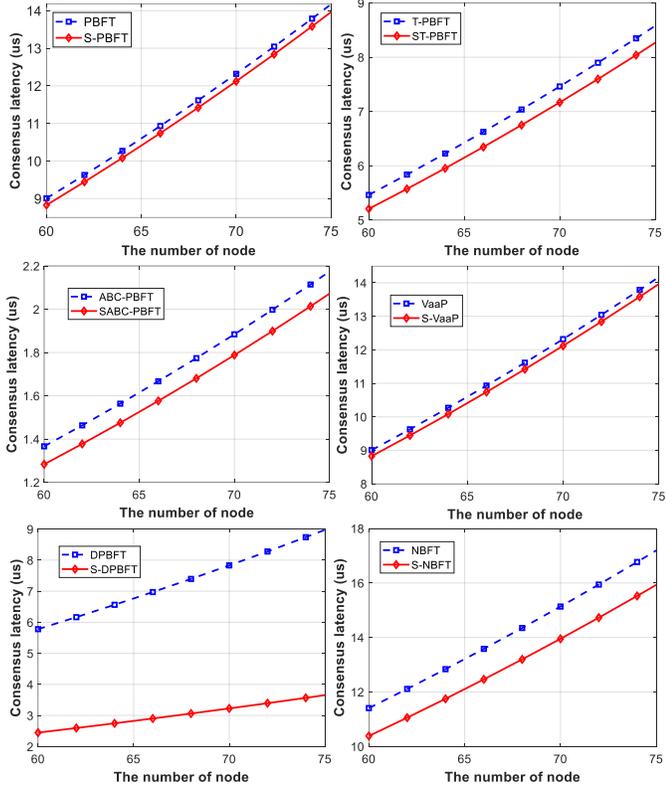

**Fig. 9.** Consensus latency of PBFT-like CMs under TD multiplexing.

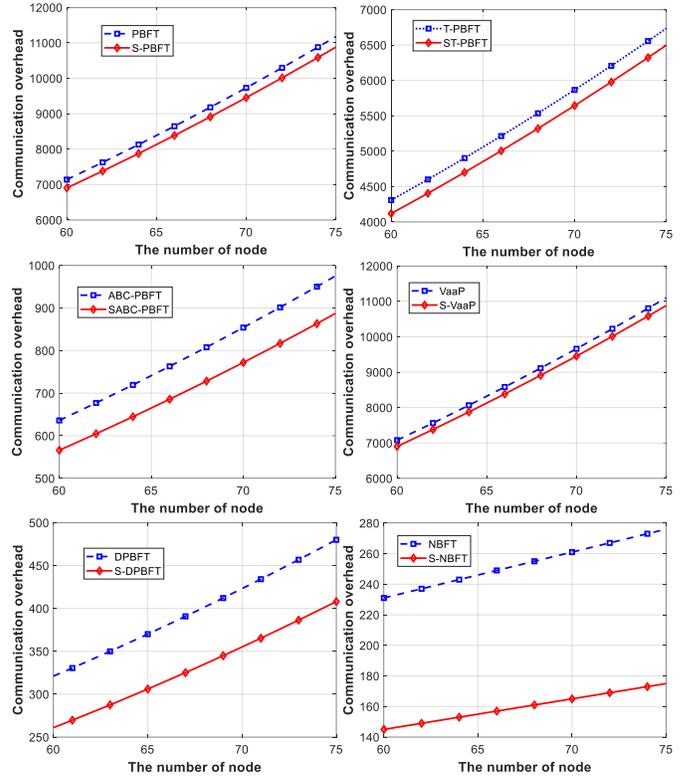

**Fig. 10.** Communication overhead of PBFT-like CMs.

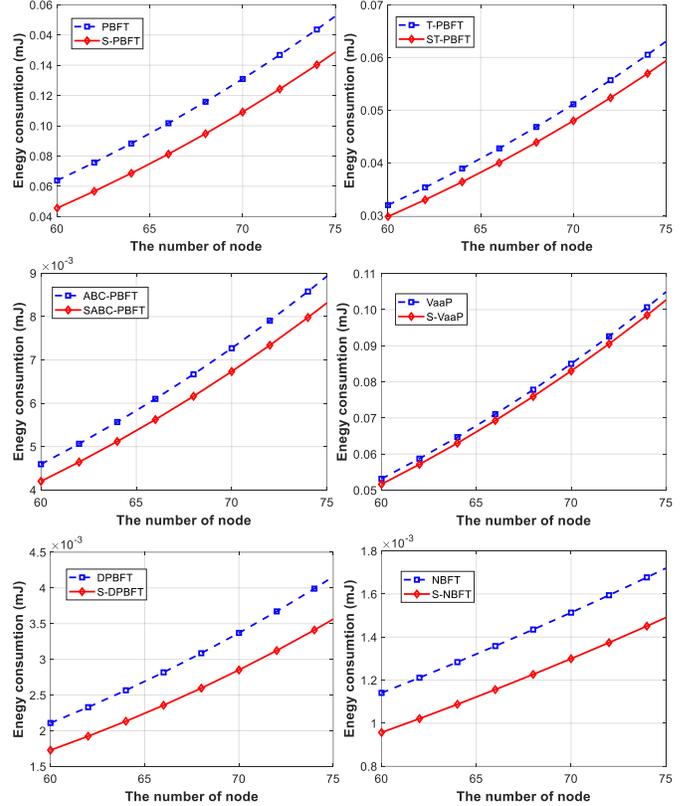

**Fig. 11.** Energy consumption of PBFT-like CMs.

Second, we use Fig. 8-9 to show the **consensus latency** of these 6 PFT-like CMs with TD and FD multiplexing. Since we set a fixed FB value when the *n* value in WBN is too large, a large number of SR systems will occupy a lot of spectrum resources under FD. For such cases, we need to use TD to avoid signal interference between SR systems. In the FD case, the SBC latency has a two-order magnitude advantage over the TD case. When the *n* value is 75, the average advantage of the 6 PBFT-like SBCs in terms of latency is 3.2% and 12.1% in the case of FD and TD respectively. Among them, we find that S-DPBFT and S-NBFT achieve significant latency optimization under both multiplexing. This may be that sharding further amplifies the latency advantage of SBC.

Third, Fig. 10 shows the **communication overhead** for the 6 PBFT-like CMs. Simulation results show that regardless of which CMs, SBCs can reduce communication overhead after the deployment of the SR system. The communication overhead of VaaP seems to be consistent with that of PBFT, because the rollback mechanism of VaaP can only improve the parallel efficiency of multiple broadcasts, but cannot reduce the communication overhead. Other CMs save communication overhead by reducing consensus nodes or using the sharding method, etc. In addition, similar to the simulation results of latency, S-DPBFT and S-NBFT achieve the greatest optimization effect in terms of communication overhead. It may be that sharding provides greater optimization space for the SBC since the SR system can be deployed separately in each shard. When *n*=75, the 6 PBFT-like SBCs saved an average of 12.2% communication overhead.

Finally, SBC also has a relatively notable optimization effect in terms of **energy consumption**. When the number of nodes is 75, the 6 PBFT-like SBCs can save energy by an average of 9.2%. And with the increase in the number of nodes, this value will increase. Like the consensus success rate, energy consumption is one of the most important performance indicators in WBNs, which determines whether the WBN can work sustainably, so its optimization has great significance. Meanwhile, except for the two sharding CMs, the energy consumption of ABC-PBFT is also very low compared with that of PBFT, because only some nodes in the network are selected to participate in the consensus. In addition, in this indicator, S-DPBFT and S-NBFT also achieve the most significant energy optimization. The reason for this should be the same as the previous two properties.

*B. Performance Evaluation of RAFT-like CM*

First, we also analyze the key that determines whether the WBN can reach consistency, namely the RAFT-like **consensus success rate**. Unlike PBFT-like CMs, the consensus success rate of RAFT-like presents a fluctuating upward trend with the increase of the number of nodes, which is consistent with [14], [21]. Surprisingly, RAFT has a higher consensus success rate than other improvement schemes. This reason is that KRAFT takes some followers as leaders, thus reducing the number of nodes participating in consensus; Two-hop nodes in TH-RAFT are more susceptible to unstable wireless links. VSSB-RAFT divides the network into a region of 10 nodes, similar to sharding. Whatever, SBCs enhanced by multipath gain have a higher consensus success rate than the original CMs. When $n=4$, $P_s=0.8$, and $P_e=0.9$, 4 RAFT-like SBCs improved the consensus success rate by an average of 5.8%. Compared with PBFT-like CMs, the optimization effect is not very evident, because the consensus success rate of RAFT-like CMs is already high.

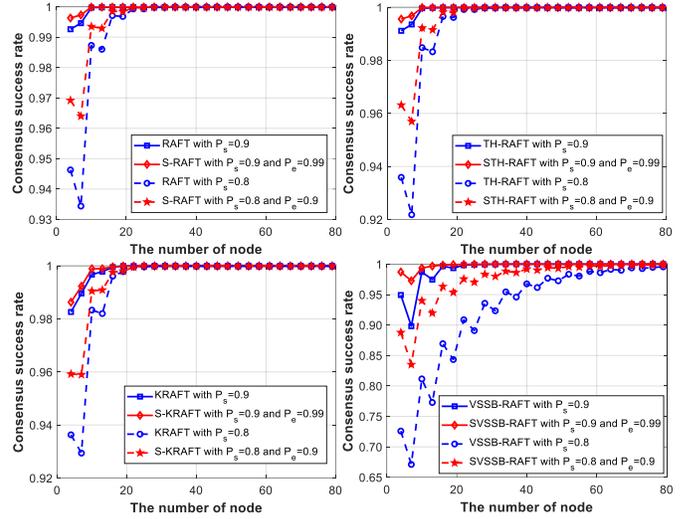

**Fig. 12.** Consensus rate of RAFT-like CMs.

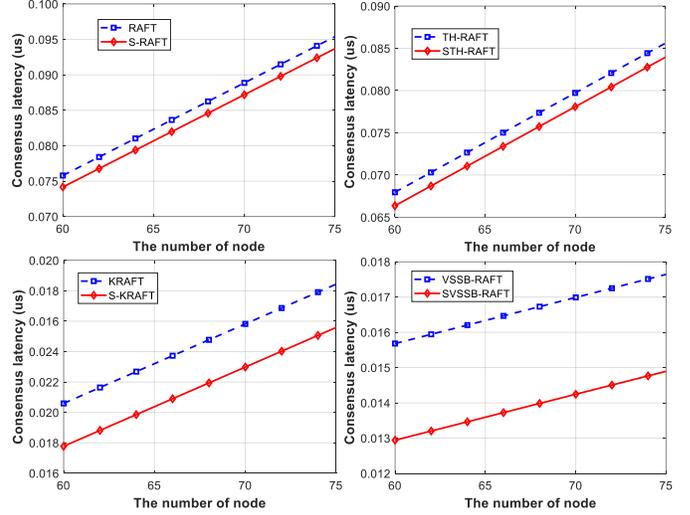

**Fig. 13.** Consensus latency of RAFT-like CMs under FD multiplexing.

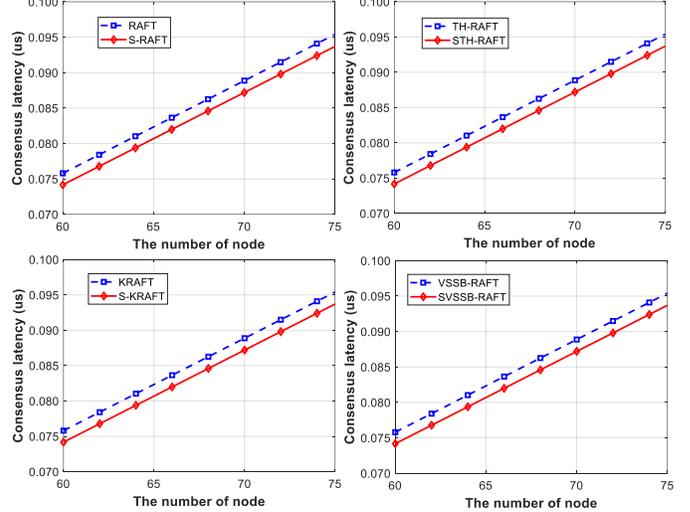

**Fig. 14.** Consensus latency of RAFT-like CMs under TD multiplexing.

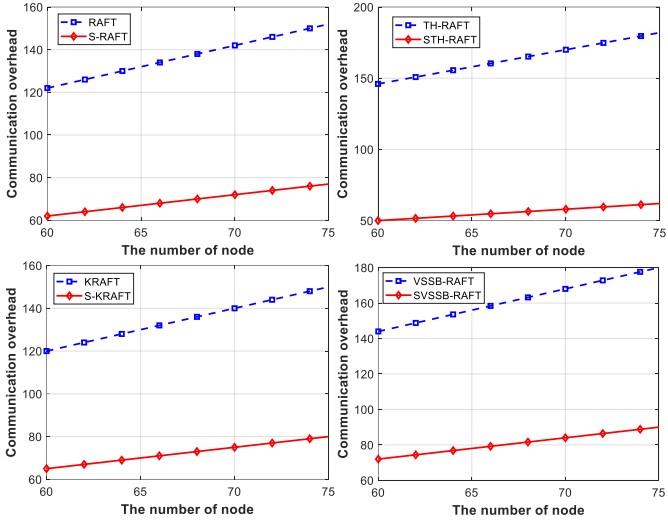

**Fig. 15.** Communication overhead of RAFT-like CMs.

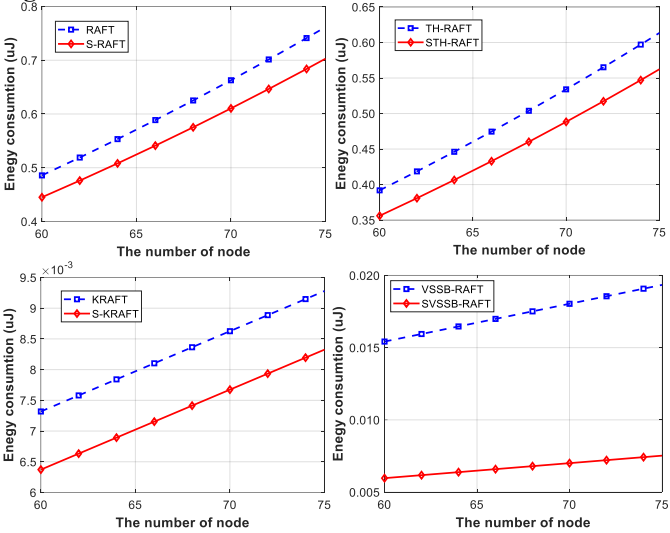

**Fig. 16.** Energy consumption of RAFT-like CMs.

Second, we analyze the **consensus latency** of RAFT-like CMs in FD and TD cases, as shown in Fig. 13-14. Similarly, TD is for too many nodes situation. RAFT consensus latency is the same in both multiplexing modes, because there are no parallel broadcast processes. This is consistent with Section V-B-2. The other three improved RAFT CMs have longer latency in the TD case due to their parallel broadcasts. Fortunately, the latency of these three CMs only degenerated to the same level as RAFT in the TD case, because with this multiplexing, the consensus latency is only related to the number of nodes [16]. This also shows that RAFT-like CMs are more suitable for TD multiplexing since it does not occupy too many spectrum resources and have a lower upper limit of latency. Unlike PBFT-like SBCs, RAFT-like SBCs have a better latency optimization effect of 6.9% in the FD case, and only 3.2% in the TD case when $n$ is 75. The reason is that TD degrades each CM to the RAFT level in consensus latency.

Additionally, SVSSB-RAFT with FD has a more significant latency optimization effect, because it uses a similar sharding method to divide the network into multiple regions.

Third, the **communication overhead** of RAFT-like CMs is evaluated in Fig. 15. TH-RAFT and VSSB-RAFT increase the communication overhead compared with RAFT. For TH-RAFT, there are partial two-hop nodes, and more communication processes are needed to complete the consensus. For VSSB-RAFT, it uses a sharding-like method, so a committee consensus is needed. Moreover, consistent with the analysis in Section V-B-3, RAFT-like SBCs can achieve an optimization effect of about 50% in terms of communication overhead.

Finally, we have evaluated the progress made in **energy consumption** by RAFT-like CMs. All three optimized RAFT CMs consume less energy than the original RAFT due to lower communication loads. In addition, RAFT-like SBCs have achieved a great energy-saving effect. When the number of nodes equals 75, they reduce energy consumption by an average of 23.7%. The reason that the energy consumption is not optimized well as the communication overhead is that the *uplink* communication phase, which is replaced by passive communication in RAFT-like CMs, is not a broadcast process.

In general, RAFT-like CMs have higher consensus success rate, lower consensus latency, communication overhead, and energy consumption than PBFT-like CMs, because the former has a more simplified negotiation process.

## VII. CONCLUSION

Inspired by SR systems, we propose SBC based on cognitive backscatter communications. This new concept can effectively solve the problem of unstable communication links and high energy consumption in WBNs. To demonstrate its universality, we select 10 SOTA CMs as representatives, including 6 PBFT-like and 4 RAFT-like CMs. Meanwhile, for different spectrum resource situations, we also provide FD and TD multiplexing technologies to avoid interference in the same frequency. In particular, we present design details for S-PBFT and S-RAFT, as well as simulations for all SBCs. The results show that PBFT-like SBCs have achieved an average consensus success rate of 54.1% and reduced energy consumption by 9.2%; RAFT-like SBCs have improved consensus success rates by an average of 5.8% and reduced energy consumption by 23.7%. In addition, SBC also optimizes consensus latency and communication overhead.

In general, SBC provides a new paradigm for WBNs, which can make full use of the spectrum resources, and is suitable for almost all vote-style CMs relying on multi-communication negotiations. In the future, we will further study symbiosis-oriented blockchain consensus, sharding, and node deployment.

## APPENDIX

During the *commit* phase of S-PBFT, $n-1$ replicas act as receivers with $n-3$ active communications and 2 passive communications. To meet the fault tolerance requirement of 1/3, this $n-1$ communication allows a maximum of $b$ failures. The threshold of $b$ can be further divided into three cases: 1) at most $b-2$ active communications; 2) 1 passive communication and *b*-1 active communications; 3) 2 passive communications and *b*-2 active communications.

For these three cases, their probabilities are (32)-(34) respectively. Then, sum these three probabilities to get (10).

$$\sum_{k=0}^{b-2} C_{n-3}^{k} \left(1-P_e\right)^k P_e^{(n-3-k)}, \quad (32)$$

$$C_{n-3}^{b-1} \left(1-P_e\right)^{b-1} P_e^{(n-b-2)} C_2^1 P_s (1-P_s), \quad (33)$$

$$C_{n-3}^{b-2} \left(1-P_e\right)^{b-2} P_e^{(n-b-3)} P_s^2. \quad (34)$$

Furthermore, at this phase, the number of nodes that do not receive more than 2/3 *commit* messages cannot exceed *b*. These *b* nodes can be divided into two cases: 1) at most *b*-1 replicas; 2) the primary node and *b*-1 replicas.

For these two cases, their probabilities are (35)-(36) respectively. Then, sum these three probabilities to get (11).

$$\sum_{l=0}^{b-1} C_{n-1}^{l} \left(1-P_{\text{Re}}\right)^l P_{\text{Re}}^{(n-1-l)}, \quad (35)$$

$$C_{n-1}^{b-1} \left(1-P_{\text{Re}}\right)^{b-1} P_{\text{Re}}^{(n-b)} P_{PN}. \quad (36)$$